\documentclass[a4paper]{article}
\usepackage[utf8]{inputenc}
\usepackage{graphicx}
\usepackage{xcolor}
\usepackage{multicol}
\usepackage[ruled,vlined]{algorithm2e}
\usepackage{float}
\usepackage{hyperref}

\title{Helsinki Deblur Challenge 2021: \\description of photographic data}
\author{Markus Juvonen, Samuli Siltanen and Fernando Silva de Moura}
\date{March 2021}

\begin{document}

\maketitle

\begin{abstract}
    The photographic dataset collected for the Helsinki Deblur Challenge 2021 (HDC2021) contains pairs of images taken by two identical cameras of the same target but with different conditions. One camera is always in focus and produces sharp and low noise images the other camera produces blurred and noisy images as it is gradually more and more out of focus and has a higher ISO setting.  
    Even though the dataset was designed and captured with the HDC2021 in mind it can be used for any testing and benchmarking of image deblurring algorithms. The data is available here: \href{https://doi.org/10.5281/zenodo.4772280}{\color{blue}{https://doi.org/10.5281/zenodo.477228}}
    
\end{abstract}

\tableofcontents

\section{Introduction}

This document reports the acquisition, structure and properties of a digital photographic dataset collected at the Industrial Mathematics Laboratory of the Department of Mathematics and Statistics of University of Helsinki, Finland. This dataset was primarily designed and captured to be used for the Helsinki Deblur Challenge 2021 but it can be used for any testing and benchmarking purposes of image deblurring algorithms. 

The dataset contains pairs of photographs of random strings of text. The pairs of images are taken using two identical cameras of the same text file but with different focus conditions. One of them produces sharp images while the other camera is purposely misfocused in increasing and controlled amounts to produce blurred images of various levels. Why strings of text as targets? This is to enable quantitative measurement of deconvolution quality using an Optical Characted Recognition (OCR) algorithm in the Helsinki Deblur Challenge 2021.

However, the openly available sharp-blurred image pairs shot under controlled empirical conditions can be used for many research purposes.

The data can be found here: \href{https://doi.org/10.5281/zenodo.4772280}{\color{blue}{https://doi.org/10.5281/zenodo.477228}}

\section{Contents of the Dataset}





\begin{figure}[H]
    \begin{picture}(320,575)
    \put(90,0){\includegraphics[width=7.05cm]{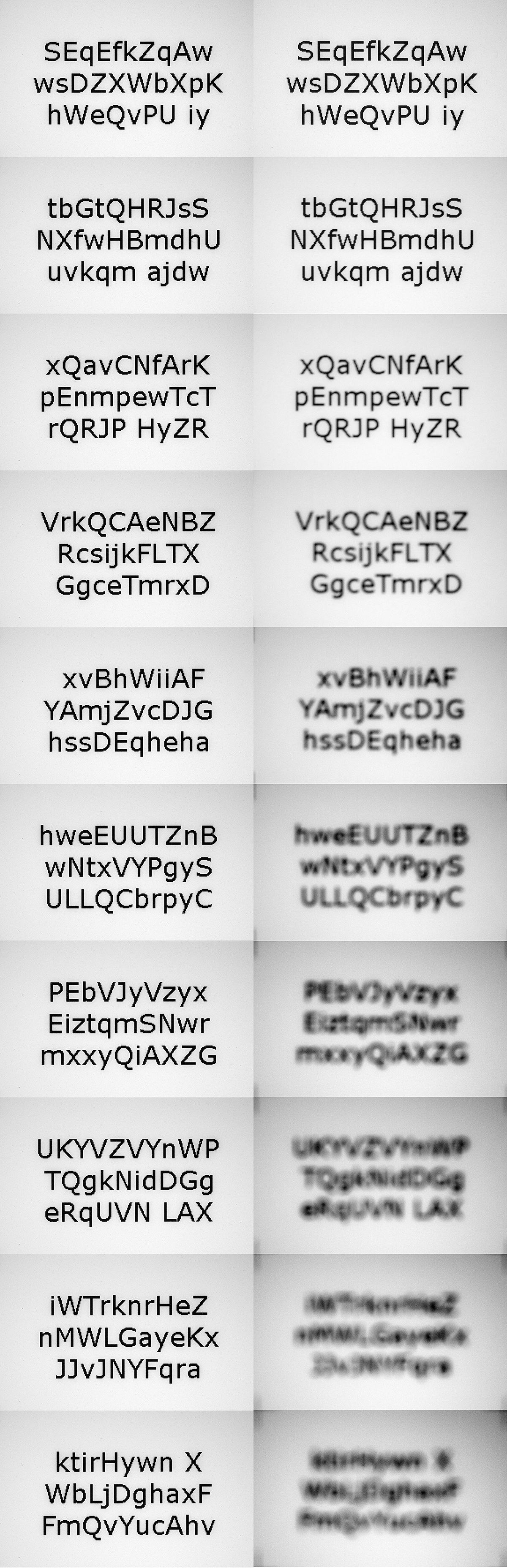}}
    \put(111,624){Sharp image}
    \put(207,624){Blurred image}
    \put(10,580){Focus step 0}
    \put(10,517.8){Focus step 1}
    \put(10,455.5){Focus step 2}
    \put(10,393.3){Focus step 3}
    \put(10,331.1){Focus step 4}
    \put(10,268.9){Focus step 5}
    \put(10,206.7){Focus step 6}
    \put(10,144.4){Focus step 7}
    \put(10,82.2){Focus step 8}
    \put(10,20){Focus step 9}
    \end{picture}    \caption{A sample set of 10 image pairs. One pair for each amount of blur. }
    \label{fig:sample_set}
\end{figure}

\begin{figure}[H]
    \begin{picture}(320,295)
    \put(10,0){\includegraphics[width=13cm]{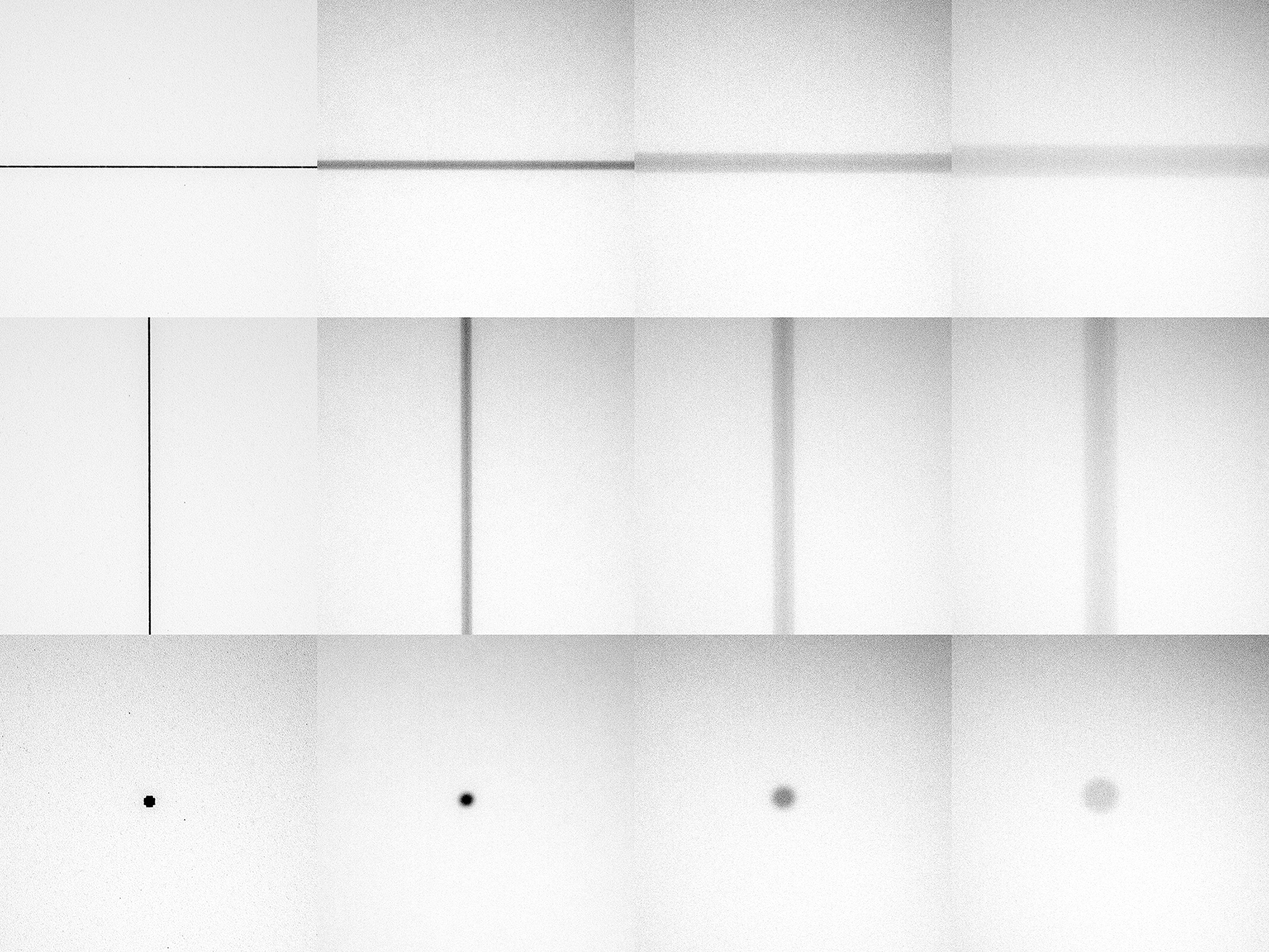}}
    \put(30,290){Focus step 0}
    \put(119.7,290){Focus step 3}
    \put(212.3,290){Focus step 6}
    \put(305,290){Focus step 9}
    \end{picture}
    \caption{Line and point spread functions for focus steps 0, 3, 6 and 9. }
    \label{fig:psf_sample_set}
\end{figure}
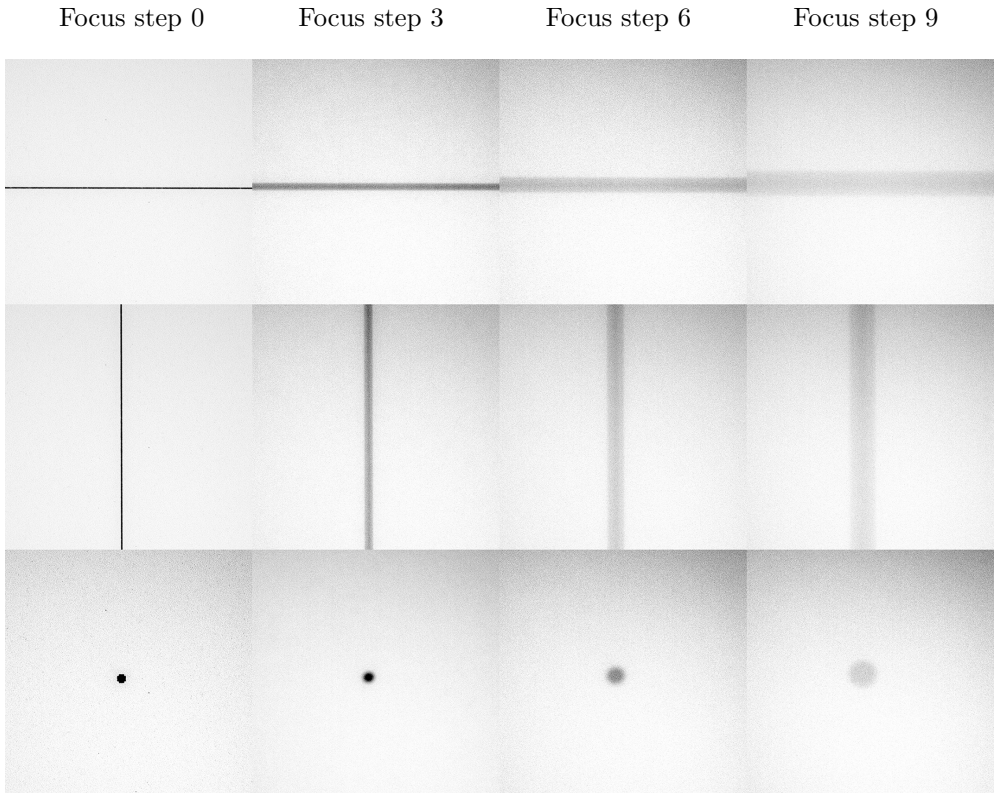

The dataset consists of 2000 image pairs. The two image files considered a pair are pictures taken with the two cameras of the same text target (see section \ref{sec:methods} for detailed description). One picture is well focused and the other has some level of blur (Figure \ref{fig:sample_set}). The blurring results from adjusting one of the camera's  focus plane step by step away from the plane where the text target is located. Two different fonts were used, Verdana and Times, so there are 1000 pairs for each font. Each image is accompanied by a text file with the transcription of that particular text target. For each step of increased blur, additional images of horizontal and vertical line spread functions, as well as a point spread function are provided (Figure \ref{fig:psf_sample_set}).

On the Zenodo repository (\href{https://doi.org/10.5281/zenodo.4772280}{\color{blue}{https://doi.org/10.5281/zenodo.477228}}) the images are split into 10 separate zip files according to the level of blur. Each one of the zip files contains two folders, one for each of the different fonts used (Verdana and Times). In both font folders you find one folder named CAM1 (Camera 1) with the sharp images and one named CAM2 (Camera 2) with the blurred images. Each one of the CAM folders holds 100 images of the text targets and 3 images of technical targets for estimating the point spread function (PSF). Each one of the text target images is accompanied by a text file (same file name with .txt extension) containing the correct transcription of that particular text target.



\section{Materials and equipment}\label{sec:methods}

\begin{figure}[H]
    \centering
    \includegraphics[width=12cm]{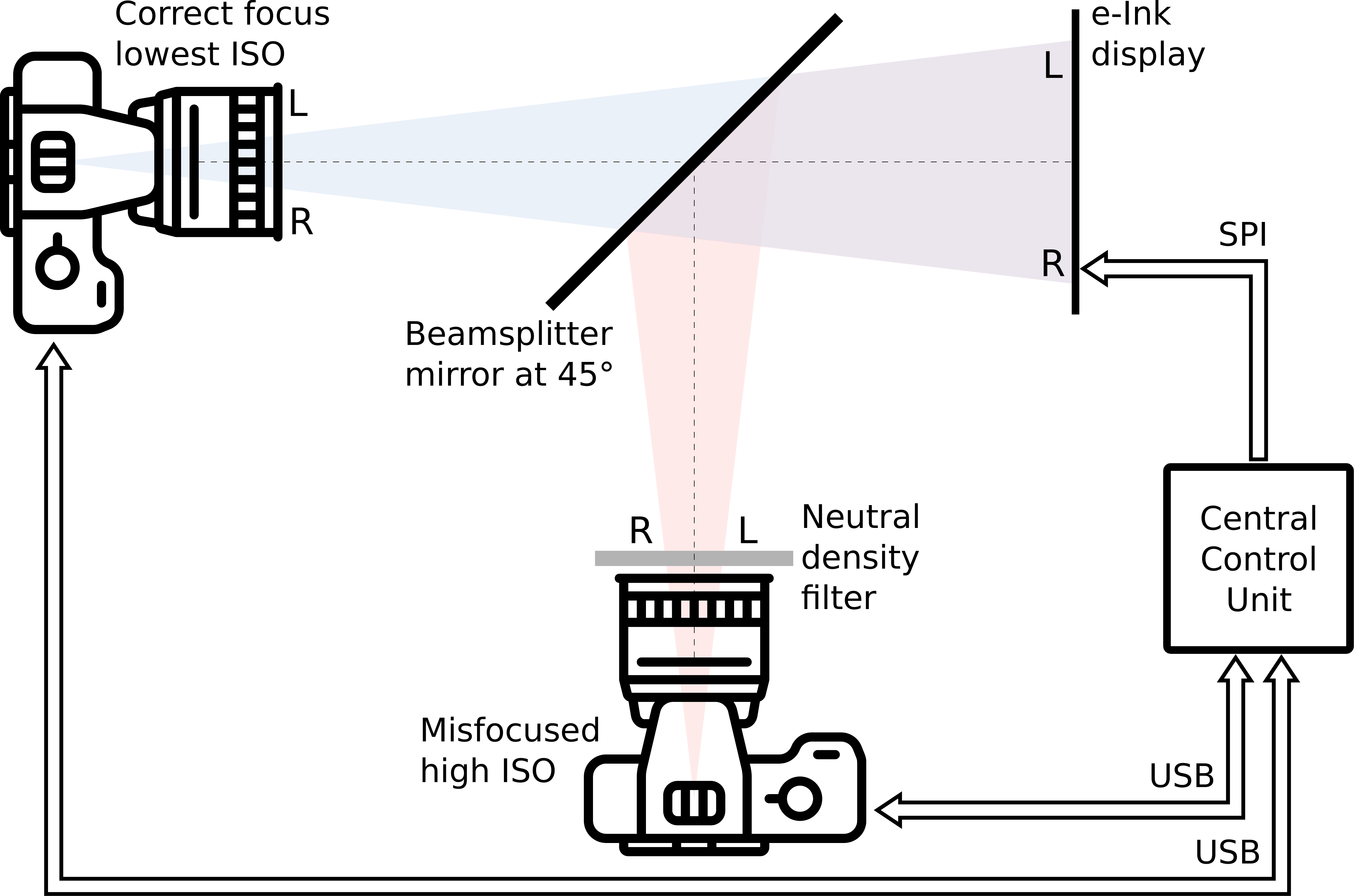}
    \caption{Diagram of the experiment setup. The mirror is half-transparent. Note that the image recorded by Camera 2 is flipped due to the mirror. }
    \label{fig:Cameras}
\end{figure}

The experimental setup was built on two Thorlabs breadboards that were connected through dovetail optical rails. The cameras and the beamsplitter mirror were attached to the larger breadboard and the E-ink display to the smaller one. One Olight X6 Marauder LED light on medium setting (approx. 1200 lumen) was used to illuminate the display in addition to the ambient lighting in the room. 
Here are some more detailed descriptions of the most important parts of the measurement set up as seen in Figure \ref{fig:Cameras}.

\subsection{Beamsplitter Mirror}

We used a 15cm x 15cm sized and 2mm thick 50T-50R beamsplitter mirror from Optical Mirror company. The mirror is fixed at a 45 degree angle relative to both cameras as seen in Figure \ref{fig:Cameras}. Ideally a beamsplitter mirror would let us take exactly the same image with both cameras one of them just being a mirror image. But as it is a real mirror the ideal reflectance and transmission of 50 percent are not realistic. The manufacturer specifies the reflectance of the mirror at 45 degrees at 50\% with a tolerance of +3\%, -7\%. The typical overall transmission value at 45 degrees is 46\%. More detailed technical specifications of the used 50T-50R mirror can be found on \href{https://opticalmirror.com/beamsplitter/}{\color{blue}{https://opticalmirror.com/beamsplitter/}}.

\subsection{Camera equipment, arrangement and settings}
We shot with two Canon EOS 5D Mark IV cameras arranged as shown in Figure \ref{fig:Cameras}. Both cameras have a Canon EF 100mm f/2.8 USM Macro lens attached. 
\begin{itemize}
    \item Camera 1 has ISO setting 100 and is well-focused throughout the measurements. The shutter speed is selected to be 1/160s and the aperture f/7.1 to have a well lit (no over- or underexposed parts) image in the lighting conditions used. 
    \item Camera 2 has a higher ISO setting of 6400 and a variable neutral-density (ND) filter to match the brightness of the image at the same shutter speed and aperture settings as camera 1. Also, it's focus will gradually get worse during the measurements.
\end{itemize} 

Camera 1 was covered up with a thick black cloth as seen in Figure \ref{fig:cam1hidden} to prevent any unwanted reflections. The fine adjustment of the ND filter was aided by comparing the pixel intensity histograms of both cameras and adjusting the filter until they matched as well as possible.

\subsection{E-ink Display}

\begin{figure}[]
    \centering
    \includegraphics[width=15cm]{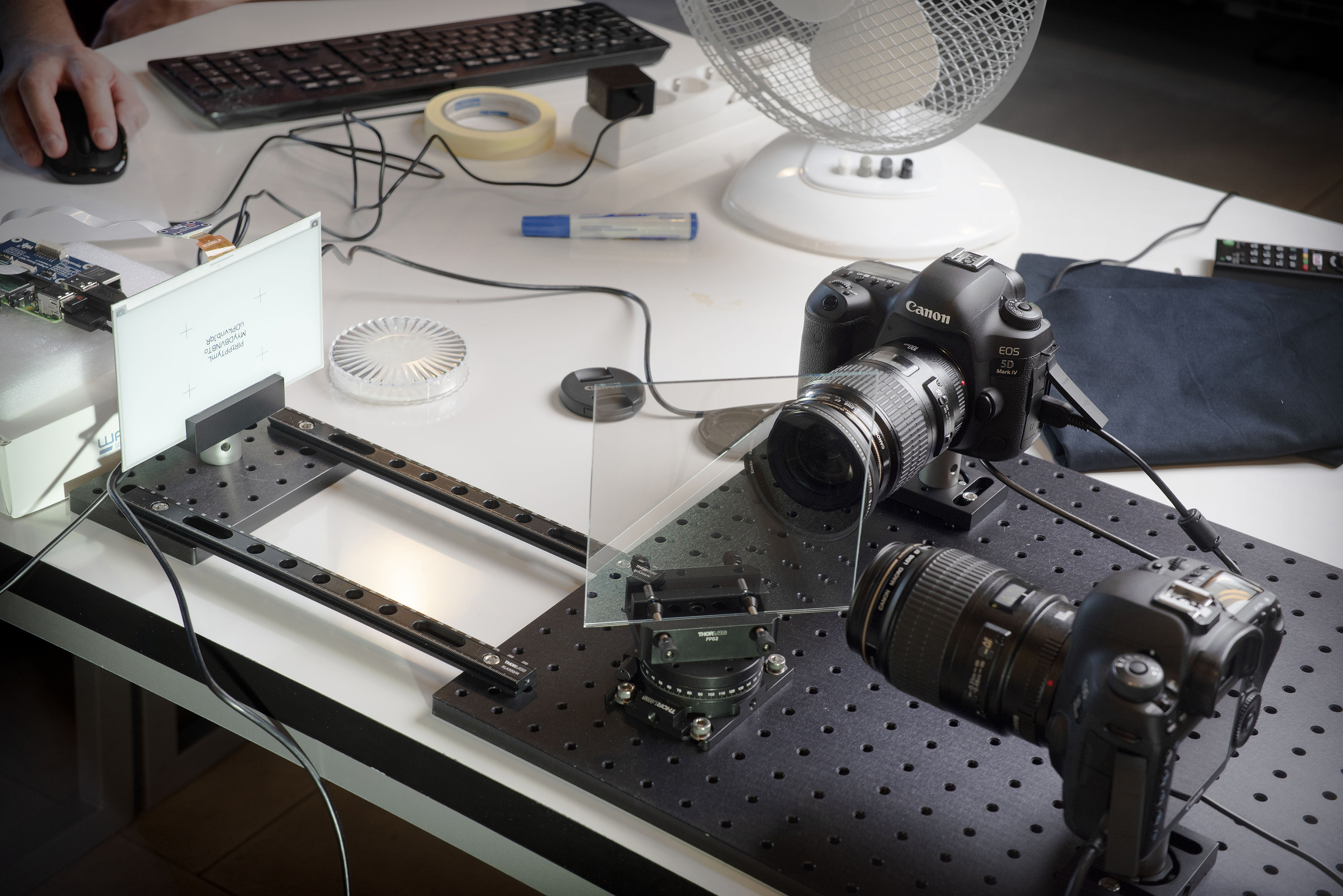}
    \caption{Image showing the cameras, the mirror and the e-ink display set up on the breadboards. The camera in the bottom right-hand corner and the beam-splitter mirror were covered with black cloth during shooting, as shown in Figure \ref{fig:cam1hidden}. }
    \label{fig:setup_test}
\end{figure}

\begin{figure}[]
    \centering
    \includegraphics[width=15cm]{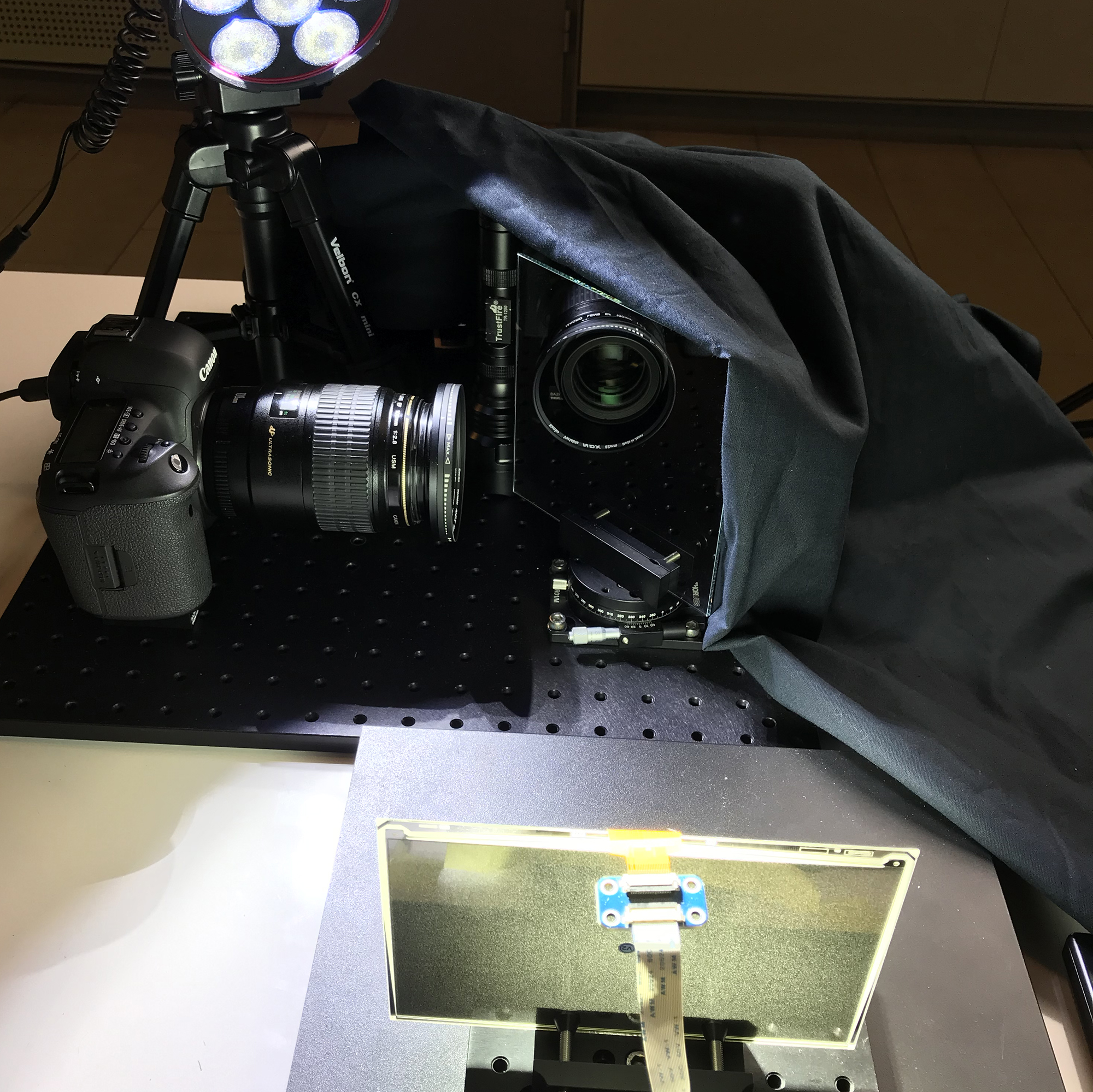}
    \caption{Image showing Camera 2, the beamsplitter mirror, Camera 1 covered by cloth and the e-ink display from behind.}
    \label{fig:cam1hidden}
\end{figure}

As the imaging target that displays the character strings we chose a E-ink screen. Using the E-ink screen removes issues with possible flickering and Moiré interference patterns that arise with more conventional displays. The display can be seen in Figures \ref{fig:setup_test} and \ref{fig:cam1hidden}. Here are more detailed specifications of the used display:

\begin{multicols}{2}
\begin{description}
    \item [Brand:] Waveshare
    \item [Model:] 800x480, 7.5inch E-Ink display
     \item [Interface:] SPI
     \item [Display size:] 163.2mm x 97.92mm
     \item [Dot pitch:] 0.205mm x 0.204mm
     \item [Resolution:] 800 x 480
     \item [Display color:] black, white
     \item [Grey scale:] 2
     \item [Viewing angle:] $>170^o$
\end{description}

\end{multicols}


\subsection{Display targets}

 The targets in the e-ink display are composed by a central 300x200 area with fiducial markers in the four corners for registration.
 
\begin{figure}[H]
    \centering
    \includegraphics[width=10cm]{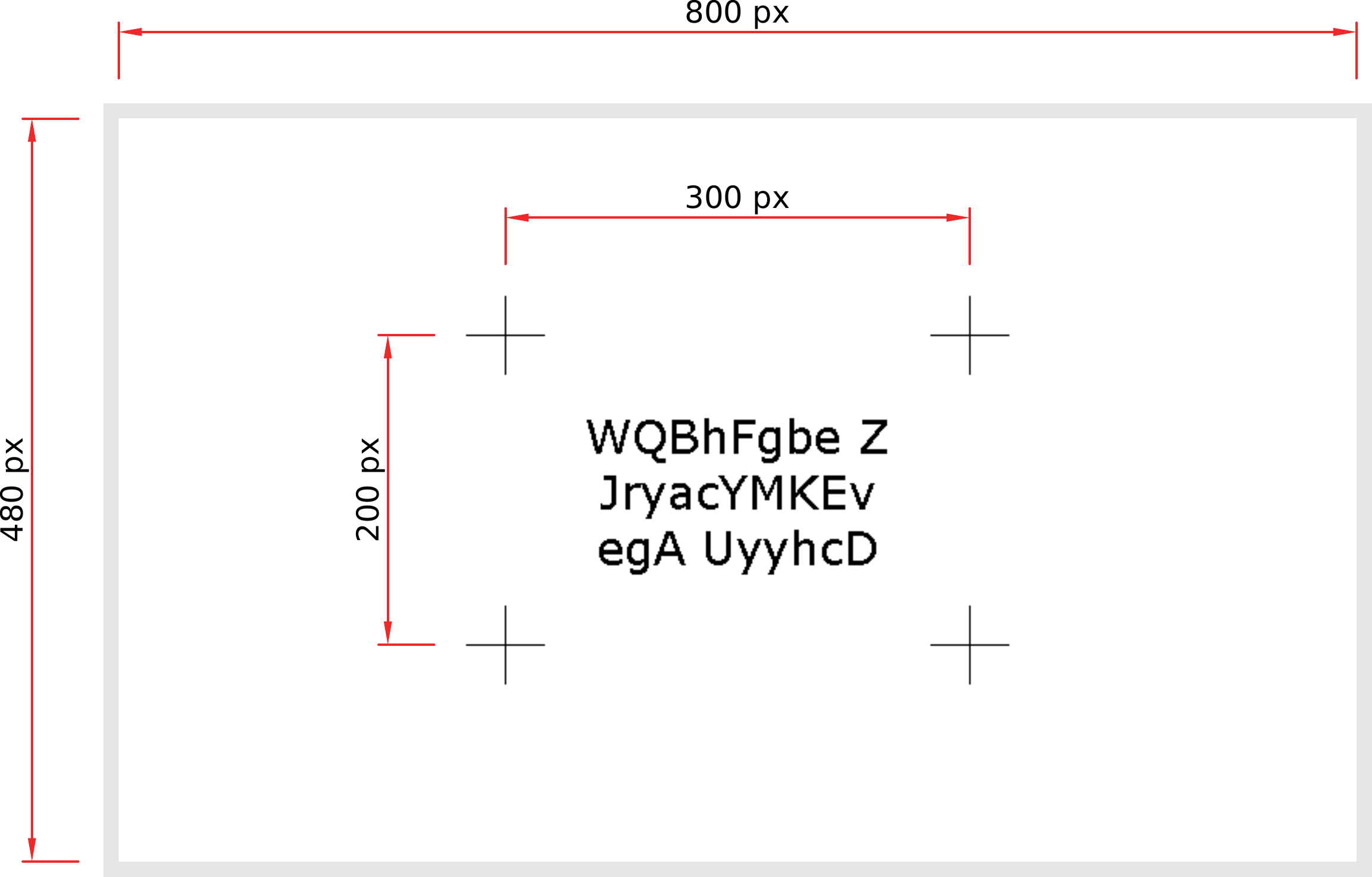}
    \caption{Target image example showing random text using 30-point Verdana font.}
    \label{fig:target}
\end{figure}

 Inside this area 3 lines of centered random strings of characters is generated with the same font type and size. See Figure \ref{fig:target} for one example.  Both the text and the fiducial marks are black while the rest of the display is white.

The random text is created with latin alphabet characters (upper and lower cases) without any diacritical marks (accents). The primary objective of the dataset is the Helsinki Deblur Challenge and, in order to reduce ambiguity in the OCR, the characters 'o' (both lower and upper cases), '0' (zero), 'l' (lowercase L), 'I' (uppercase i) were excluded. The bottom an top lines rows can contain spaces but not the central line.

\subsection{Central control unit}

The experimental apparatus is controlled by a central processing unit. The control unit is a Raspberry Pi 4 (Model B - 8 GB RAM - 256Gb SD card) running raspbian OS.

The central control, programmed in Python 3, is responsible for creating the random text, showing it in the display and commanding the cameras to take pictures . After that, the pictures will be automatically sent to the Rpi and stored.

The cameras are controled with gPhoto2 interface via USB cables while the e-display is controlled with custom software via SPI interface.

\section{Image pre-processing}

\subsection{RGB RAW to Grayscale TIFF conversion}
 The Canon .CR2 RAW images stored by the central control unit were first converted into 16-bit grayscale TIFF images using the Adobe Photoshop (21.0.3 release) image processor. 
 
 Why grayscale? As our target is a black and white e-ink display it does not make much sense to store RGB color information. Especially as we are dealing with a huge amount of rather large image files to be shared online.

\subsection{Alignment and cropping}
Next the images from Camera 2 are flipped horizontally as they were taken via the mirror. Then the first two sharp images from both cameras are aligned horizontally and vertically so that the fiducial markers line up as well as possible according to visual inspection.

The aligned images are then cropped so that the markers are no longer visible and the size of the images is greatly reduced. The final pixel size of a single image is 2360 columns and 1460 rows (original size was 6720 columns and 4480 rows). In Figure \ref{fig:sample_set} you can see that with more blur the fiducial markers start to spread into the corners of the blurred images.

Both the alignment and the cropping were done in MATLAB. 


\end{document}